\documentclass[12pt]{article}
\usepackage{a4wide}
\usepackage{epsfig}
\usepackage{amsmath}

\newlength{\absize}
\catcode`@=11
\def\citer{\@ifnextchar [{\@tempswatrue\@citexr}{\@tempswafalse\@citexr[]}}

\def\@citexr[#1]#2{\if@filesw\immediate
  \write\@auxout{\string\citation{#2}}\fi
  \def\@citea{}\@cite{\@for\@citeb:=#2\do
    {\@citea\def\@citea{--\penalty\@m}\@ifundefined
       {b@\@citeb}{{\bf ?}\@warning
       {Citation `\@citeb' on page \thepage \space undefined}}%
\hbox{\csname b@\@citeb\endcsname}}}{#1}}
\catcode`@=12

\begin{document}
  \thispagestyle{empty}
  \pagestyle{empty}
  \renewcommand{\thefootnote}{\fnsymbol{footnote}}
\newpage\normalsize
    \pagestyle{plain}
    \setlength{\baselineskip}{4ex}\par
    \setcounter{footnote}{0}
    \renewcommand{\thefootnote}{\arabic{footnote}}
\newcommand{\preprint}[1]{%
  \begin{flushright}
    \setlength{\baselineskip}{3ex} #1
  \end{flushright}}
\renewcommand{\title}[1]{%
  \begin{center}
    \LARGE #1
  \end{center}\par}
\renewcommand{\author}[1]{%
  \vspace{2ex}
  {\Large
   \begin{center}
     \setlength{\baselineskip}{3ex} #1 \par
   \end{center}}}
\renewcommand{\thanks}[1]{\footnote{#1}}
\begin{flushright}
Revised, February 18, 2006
\end{flushright}
\vskip 0.5cm

\begin{center}
{\large \bf  Quantum Anomaly Dissociation of Quasibound States
Near the Saddle-Point Ionization Limit of a Rydberg Electron in
Crossed Electric and Magnetic Fields}
\end{center}

\begin{center}
\noindent  Jian-Zu Zhang$\;^{1),\;\ast}$, Li-Ming
He$\;^{1),\;2)}$, Yun-Xia Zhu$\;^{2)}$
\end{center}
\begin{center}
1) Institute for Theoretical Physics, East China University of
Science and Technology, Box 316, Shanghai 200237, P. R. China \\
2) Department of Physics, East China University of Science and
Technology, Shanghai 200237, P. R. China

\end{center}

\vspace{1cm}


\begin{abstract}
In the combination of crossed electric and magnetic fields and the
Coulomb field of the atomic nucleus the spectrum of the Rydberg
electron in the vicinity of the Stark saddle-point are
investigated at a quantum mechanical level. The results expose a
quantum anomaly dissociation: quasibound states near and above the
saddle-point ionization limit predicted at the semi-classical
level disappear at a quantum mechanical level.

\end{abstract}

\vspace{0.4cm}

\begin{flushleft}
PACS numbers: 32.60.+i

\vspace{0.4cm}
$^{\ast}$ Corresponding author
\end{flushleft}
\clearpage
Twenty years ago Clark et al. \cite{Clar} claimed that the
combination of crossed electric and magnetic fields and the
Coulomb field of the atomic nucleus can lead to the localization
of the Rydberg electron in the vicinity of the Stark saddle-point.
When the characteristic parameter $\omega_t^2>0$ in the classical
equation of motion the electron motion is periodic in an
elliptical orbit. Such orbits give rise to electron states which
are localized above the saddle-point and whose spectrum is that of
a harmonic oscillator. Divergent hyperbolic trajectories are
obtained in the case when $\omega_t^2<0$. The classical equation
of motion includes both type of solutions. The periodic orbits are
unstable with respect to small perturbations, thus assume the
character of quasibound states. Ref.~\cite{Clar} focused attention
on the energy region near the ionization threshold for the first
time in literature. The distinctive character of this portion of
the spectrum makes it an attractive target of experimental
investigation. Their treatment is semi-classical. The
determination of the lifetimes of these states and their
associated transition moments awaits a full quantum mechanical
treatment.

The full quantum mechanical treatment of the
above-ionization-threshold spectra of atoms in crossed electric
and magnetic fields can be investigated globally and locally. In
literature there were a lot of works focused on the global aspect
of this problem, for example, see Main and Wunner \cite{MW}, Main,
Schwacke and Wunner \cite{MSW}, etc. and references there in.

Clark et al \cite{Clar} considered, classically, the local aspect
of the above system.

In this Letter we investigate the local aspect of the above system
for Clark's case \cite{Clar} at a quantum mechanical level. The
results reveal a quantum anomaly dissociation: bound states which
exist at a semi-classical level may disappear at a quantum
mechanical level. For the present example, we find that quasibound
states of the harmonic type
above the saddle-point ionization limit predicted in
Ref.~\cite{Clar} do not exist at a quantum mechanical level. This
explains the reason that non of the suggested experiments yet has
been realized.

Let the constant electric field ${\bf E}=-E{\bf i}$, and the
uniform magnetic field $\vec{B}$ aligning the $x_3$ axis. We can
choose a gauge so that the corresponding vector potential $A_i$
reads
$A_i=\frac{1}{2}\epsilon_{ij}B_i \tilde x_j$,
where $\epsilon_{ij}$ is a 2-dimensional antisymmetric unit
tensor, $\epsilon_{12}=-\epsilon_{21}=1,$
$\epsilon_{11}=\epsilon_{22}=0.$
In the combination of the crossed uniform magnetic and electric
fields, and the Coulomb field of the atomic nucleus the
Hamiltonian $H$ of the Rydberg electron, {\it globally}, reads
(the summation convention is used henceforth)
\begin{equation}
\label{Eq:H}
H=\frac{1}{2\mu} \tilde p_i^2-\frac{e^2}{\tilde
r}-\frac{1}{2}\omega_c\epsilon_{ij}\tilde p_i \tilde
x_j+\frac{1}{8}\omega_c^2 \tilde x_i^2-eE\tilde x_1,\; (i, j=1, 2)
\end{equation}
where $\tilde r=(\tilde x_1^2+\tilde x_2^2+\tilde x_3^2)^{1/2}$,
and $(\tilde x_1, \tilde x_2, \tilde x_3)$ are the coordinates of
the electron centered about the atomic nucleus. In the above $\mu$
and $-e$ are, respectively, the mass and the electric charge of
the electron; The magnetic cyclotron frequency $\omega_c=eB/\mu
c$.

A particle trap using static fields must confine the electron
about the Stark saddle-point where the net electric force
vanishes. We therefore consider, locally, the Schr\"odinger
equation in coordinates centered about the saddle point rather
than about the atomic nucleus. The coordinate of the saddle point
is $x_{10}=\sqrt{e/E}$. In this coordinate system the coordinates
of the electron are ($x_1$, $x_2$, $x_3$). The electrostatic
potential is given by
$\Phi=e/[(x_1+x_{10})^2+x_2^2+x_3^2]^{1/2}+E(x_1+x_{10})$.
A harmonic approximation of the potential in the region around the
saddle point is enough. For small $x_1$, $x_2$ and $x_3$ the
electrostatic potential is approximated by
$\Phi=-\frac{1}{e}V_c-\frac{1}{2}\omega_z^2(-2x_1^2+x_2^2+x_3^2)$,
where $V_c=-2e\sqrt{eE}$ is the energy of the classical ionization
limit in the presence of the electric field, and
$\omega_z^2=e^2/\mu x_{10}^3$ is the axial frequency. The
Hamiltonian $H$ of this system can be decomposed into a
2-dimensional Hamiltonian $H_{\bot}$ and a one-dimensional
harmonic Hamiltonian $H_z$ with the axial frequency $\omega_z$:
$H=H_{\bot}+H_z$. The 2-dimensional Hamiltonian $H_{\bot}$  is,
{\it locally}, the type of a quasi-Penning trap
\begin{equation}
\label{Eq:H-2a}
H_{\bot}=
\frac{1}{2\mu}(p_i-\frac{1}{2}\mu\omega_c\epsilon_{ij}x_j)^2
+\frac{1}{2}\mu\omega_z^2 (-2x_1^2+x_2^2)+V_c =\frac{1}{2\mu}
p_i^2 -\frac{1}{2}\omega_c\epsilon_{ij}p_i x_j+\alpha_i x_i^2+V_c,
\end{equation}
where $\alpha_1=\mu(\omega_c^2-8\omega_z^2)/8$,
$\alpha_2=\mu(\omega_c^2+4\omega_z^2)/8$. The magnetic field
should be strong enough to satisfy a condition
$\omega_c^2>8\omega_z^2$ so that $\alpha_1>0$. At the
semi-classical level the magnetic field $B$ itself enters into the
classical equation of motion. At a quantum mechanical level the
vector potential $A_i$ enters into the Schr\"odinger equation.
Comparing Eq.~(\ref{Eq:H-2a}) with the coefficients $\omega_z^2$
of $x_i$ terms in the classical equation of motion in
Ref.~\cite{Clar}, it shows that the coefficients $\alpha_i$ of
$x_i^2$ terms in the Schr\"odinger equation include more
information \cite{Wu-Yang}.

In the following discussions the starting point is the Hamiltonian
(\ref{Eq:H-2a}), that is, we shall take the Hamiltonian
(\ref{Eq:H-2a}) as the {\it definition} of the model of the ({\it
local}) quasi-Penning trap without making further reference to the
original ({\it global}) Hamiltonian (\ref{Eq:H}).

This system is unlike the case in Ref.~\citer{Baxt,JZZ04}. Because
of lacking symmetry in the above crossed electric and magnetic
fields, the situation of this system is involved. We find that
this system is solved by the following ansatz. We define the
canonical variables $X_{\eta}$ and $P_{\eta}$ $(\eta=a, b)$ as
\begin{eqnarray}
\label{Eq:X-P}
X_a&\equiv&\sqrt{\mu\Omega_1/2\omega_1}x_1-
\sqrt{1/2\mu\Omega_1\omega_1}p_2,\;
X_b\equiv\sqrt{\mu\Omega_2/2\omega_2}x_1+
\sqrt{1/2\mu\Omega_2\omega_2}p_2,
\nonumber \\
P_a&\equiv&\sqrt{\omega_c\omega_1/2\mu\Omega_2(\omega_2-\omega_1)}p_1+
\sqrt{\mu\Omega_2\omega_c\omega_1/2(\omega_2-\omega_1)}x_2, \;
\nonumber \\
P_b&\equiv&\sqrt{\omega_c\omega_2/2\mu\Omega_1(\omega_2-\omega_1)}p_1-
\sqrt{\mu\Omega_1\omega_c\omega_2/2(\omega_2-\omega_1)}x_2,
\end{eqnarray}
In the above the parameters $\Omega_{1,2}$ and $\omega_{1,2}$ are,
respectively, defined as
\begin{eqnarray}
\label{Eq:omega-1}
\Omega_{1,2}&\equiv&\{\pm (\alpha_2-\alpha_1)+
[(\alpha_2-\alpha_1)^2+
\mu\omega_c^2(\alpha_1+\alpha_2)]^{1/2}\}/\mu\omega_c,
\nonumber\\
\omega_{1,2}&\equiv&
(\Omega_1+\Omega_2)(2\Omega_{1,2}\mp\omega_c)/4\Omega_{1,2}.
\end{eqnarray}
The above definitions give that $\Omega_{1,2}>0$, $\omega_2>0$.
From
$\omega_2-\omega_1=\omega_c(\Omega_1+\Omega_2)^2/4\Omega_1\Omega_2>0$,
it follows that $\omega_2/\omega_1=
\Omega_1(2\Omega_2+\omega_c)/\Omega_2(2\Omega_1-\omega_c)>1$,
which shows $(2\Omega_1-\omega_c)>0$, thus $\omega_1>0$.
These results confirm that the definitions of $X_{\eta}$ and
$P_{\eta}$ are meaningful. Furthermore, the canonical variables
$X_{\eta}$ and $P_{\eta}$ satisfy
$[X_{\eta}, P_{\rho}]=i\hbar\delta_{\eta\rho}$ $(\eta, \rho=a, b)$
and $[X_a, X_b]=[P_a, P_b]=0$,
which show that modes $a$ and $b$ are fully decoupled at the
quantum mechanical level. Finally, we define the parameters
$\omega_{a,b}^2$ as
\begin{equation}
\label{Eq:omega-a,b}
\omega_{a,b}^2\equiv\Omega_{1,2}\omega_{1,2}
(\omega_c\mp2\Omega_{2,1})/(\Omega_1+\Omega_2).
\end{equation}
From Eqs.~(\ref{Eq:X-P})-(\ref{Eq:omega-a,b}) it follows that the
Hamiltonian $H_{\bot}$ in Eq.~(\ref{Eq:H-2a}) decouples into two
modes
\begin{eqnarray}
\label{Eq:H-2c}
H_{\bot}&=&H_a+H_b+V_c,
\nonumber\\
H_{a,b}&=&\frac{1}{2}P_{a,b}^2\mp\frac{1}{2}\omega_{a,b}^2
X_{a,b}^2.
\end{eqnarray}
Eq.~(\ref{Eq:omega-a,b}) shows that $\omega_b^2>0$. The mode $b$
is a harmonic oscillator with the unit mass. It is worth noting
that
\begin{eqnarray}
\label{Eq:omega-a}
\omega_c-2\Omega_2=\{(\omega_c^2+3\omega_z^2)
-[(\omega_c^2+3\omega_z^2)^2-8\omega_c^2\omega_z^2
]^{1/2}\}/\omega_c>0,
\end{eqnarray}
from which Eq.~(\ref{Eq:omega-a,b}) also gives that
$\omega_a^2>0$. The possibility of $\omega_a^2$ can't be changed
through tuning external parameters like the magnetic field $B$
and/or the electric field $E$. The Schr\"odinger equation of the
mode $a$ reads
\begin{eqnarray}
\label{Eq:H-a}
i\hbar \frac{\partial}{\partial
t}\psi_a(X_a,t)=\Big(\frac{1}{2}P_a^2-\frac{1}{2}\omega_{a}^2
X_a^2\Big)\psi_a(X_a,t).
\end{eqnarray}

In the above the system (\ref{Eq:H-2a}) is solved exactly.

At the quantum mechanical level the normalization conditions of
wave functions of bound states are a key point. The minus sign of
the $X_a^2$ term in Eq.~(\ref{Eq:H-a}) elucidates that normalized
wave functions $\psi_a(X_a,t)$ of bound states of the mode $a$ do
not exist, thus for the whole system normalized wave functions
$\Psi(X_a, X_b,t)=\psi_a(X_a,t)\psi_b(X_b,t)$ of bound states do
not exist either. Thus it is impossible to locate an electronic
state. This observation reveals the phenomenon of the quantum
anomaly dissociation that quasibound states predicted in
Ref.~\cite{Clar} do {\it not} exist at the quantum mechanical
level.

{\it Discussions} - (i) The conclusion about the quantum anomaly
dissociation applies only, {\it locally}, to the Hamiltonian
(\ref{Eq:H-2a}) of the system in the region above the saddle
point.

{\it Globally}, the Hamiltonian of the system is Eq.~(\ref{Eq:H}).
Main and Wunner \cite{MW}, Main, Schwacke and Wunner \cite{MSW} et
al. performed, full quantum-mechanically, numerical calculation of
the system (\ref{Eq:H}). The results showed the existence of quite
a few bound states near or above the ionization threshold. The
features of these bound states are different from the bound states
predicted by Clark et al. \cite{Clar}. For Clark's case the energy
spectrum associated with the periodic elliptical orbits is the
type of a harmonic oscillator.

(ii) Glas, Mosel and Zint showed that \cite{GMZ} in the cranked
oscillator model when the parameters satisfy certain conditions
the square of frequencies of the two decoupled modes are positive.
The bound states exist. The situation of the Hamiltonian
(\ref{Eq:H-2a}) is different from the cranked oscillator. All the
parameters of the Hamiltonian (\ref{Eq:H-2a}) depend on the
external magnetic field $B$ and/or the electric field $E$. Their
relations are fixed. Eq.~(\ref{Eq:omega-a}) shows that the sign of
$\omega_{a}^2$ and in turn the minus sign of the term $X_a^2$ in
Eq.~(\ref{Eq:H-a}) can't be changed through tuning the external
magnetic field $B$ and/or the electric field $E$. Thus at the
quantum mechanical level the bound states of the harmonic type
corresponding to the classical periodic elliptical orbit predicted
by Clark et al. \cite{Clar} disappear exactly.

Up to now the quantum anomaly dissociation exposed in the
Hamiltonian (\ref{Eq:H-2a}) is the only example. At the present
the clarification of general conditions leading to such a
phenomenon is an open issue. Studies on this subject are important
for experimental atomic physics which are based on the
semi-classical treatment.

{\it Note in revised version} - Since submitting this paper,
Connerade group has reached the same conclusion experimentally as
the one in this paper \cite{Conn}. However, they do find the
states near the minimum of the outer well. They have lifetimes
which seem to fit tunnelling rate between the two wells. The
features of these states are different from the ones predicted by
Clark et al. \cite{Clar}.

\vspace{0.4cm}

The author would like to thank J.-P. Connerade for communication.
This work has been supported by the Natural Science Foundation of
China under the grant number 10575037 and by the Shanghai
Education Development Foundation.

\end{document}